\documentclass[aps,prb,twocolumn,showpacs]{revtex4}
\usepackage{amsmath}
\usepackage{xspace}
\usepackage{graphicx}
\newcommand{\TENS}[1]{\ensuremath{\underline{\underline{#1}}}\xspace}
\newcommand{\BiTe}{{Bi$_2$Te$_3$ }}

\newcommand{\SigRatio}{{\ensuremath{\sigma_{xx}/\sigma_{zz}} }}
\newlength{\PICTww} \PICTww=.98\textwidth
\begin{document}

\title{$Bi_2Te_3$: Implications of the rhombohedral k-space texture on
  the evaluation of the in-plane/out-of-plane conductivity anisotropy}

\author{Peter Zahn}
\affiliation{Institut f\"ur Physik, Martin-Luther-Universit\"at Halle-Wittenberg, D-06099 Halle,
Germany}
\author{Nicki F. Hinsche}
\affiliation{Institut f\"ur Physik, Martin-Luther-Universit\"at Halle-Wittenberg, D-06099 Halle,
Germany}
\author{Bogdan Yavorsky}
\affiliation{Institut f\"ur Physik, Martin-Luther-Universit\"at Halle-Wittenberg, D-06099 Halle,
Germany}
\author{Ingrid Mertig}
\affiliation{Institut f\"ur Physik, Martin-Luther-Universit\"at Halle-Wittenberg, D-06099 Halle,
Germany}
\affiliation{Max-Planck-Institut f\"ur Mikrostrukturphysik, Weinberg 2, D-06120 Halle, Germany}

\date{\today}
\begin{abstract}
Different computational scheme for calculating surface integrals in
anisotropic Brillouin zones are compared.
The example of the transport distribution function (plasma frequency)
of the thermoelectric Material \BiTe near the band edges will be
discussed. The layered structure of the material together with the
rhombohedral symmetry causes a strong anisotropy of the transport
distribution function for the directions in the basal (in-plane) and
perpendicular to the basal plane (out-of-plane).
It is shown that a thorough reciprocal space integration is necessary
to reproduce the in-plane/out-of-plane anisotropy.
A quantitative comparison can be made at the band edges, where the
transport anisotropy is given in terms of the anisotropic mass tensor.
\end{abstract}
\pacs{71.18.+y,71.20.Gj,72.20.Pa,72.80.Jc}

\maketitle

\section{Introduction}
\newlength{\PICTw}  \PICTw=0.95\linewidth

Thermoelectric materials are nowadays widely used to convert waste
heat into electrical energy or for cooling purposes.\cite{disalvo99}
The main advantage of this conversion
process is the absence of moving parts or liquid or gaseous
components. The efficiency of the process is rather low and is mainly
determined by the dimensionless parameter
$$
ZT=\frac{\sigma S^2}{\kappa} \quad ,
$$
called figure of merit. It is determined by the specific conductivity
$\sigma$, the thermopower $S$, the absolute temperature $T$,
and the thermal conductivity $\kappa$, comprising the electronic and
lattice parts $\kappa_e$ and $\kappa_L$.

New experimental techniques allow for the preparation of
nanostructered and low-dimensional thermoelectric devices which are
supposed to possess larger ZT values.\cite{hicks93,hicks93a} 
Values up to 2.4 are reported for multilayered
Bi$_2$Te$_3$/Sb$_2$Te$_3$ systems.\cite{rama99,rama01}
A microscopic understanding of these effects can be obtained by
calculating the conductivity and powerfactor in the
semi-classical limit exploiting the relaxation time approximation. 
The so-called
transport distribution function has to be determined as a function of
electron energy $E$ as integrals of surfaces of constant electron
energy in reciprocal space. \cite{mahan96}
If one considers an anisotropic material,
like \BiTe, these quantities show a strong directional dependence,
which can support the enhancement of figure of merit in thermoelectric
heterostructures.
The calculation of these isoenergetic surface integrals requires a thorough
integration in k space. The implications of different integration
scheme will be demonstrated. To this end the paper is organized as
follows.
We start with a derivation
of the transport anisotropy near the band edges as a function of
the effective mass tensor. This provides benchmark numbers at certain
energies to check
the numerical integration schemes. 
The systematic deviations of the integration schemes are related to
the anisotropic structure of the reciprocal space of the rhombohedral
lattice and the anisotropy of the band edges concerning the effective
masses are discussed afterwards.
At the end, a simplified analytical model for a two-dimensional
integration will be discussed to show the influence 
of the rhombohedral lattice anisotropy on the transport quantities. 
Results for a free electron model show a strong dependence on the
procedure to fill the reciprocal space with tetrahedrons in a
way that the symmetry operations of the lattice can still be applied
to reduce the numerical effort.

\section{Inverse mass tensor and conductivity anisotropy at band edges}

We determined the band dispersion for \BiTe at the experimental
lattice constants with the atomic positions taken from literature.
\cite{landolt-iii-41c}
The topology of the band structure is described
elsewhere. \cite{yavorsky11}
Assuming a constant relaxation time the energy dependent matrix valued
transport distribution is defined as \cite{mahan96} 

\begin{equation}
\sigma_{\alpha\beta}(E)=
\tau \frac{e^2}{(2\pi)^3\hbar}
\int\limits_{\epsilon(k)=E}
\frac{dS}{|\vec v(k)|} v_\alpha(k) v_\beta(k)
\quad,
\label{eq:def-sigma}
\end{equation}

with $\alpha$, $\beta$ the Cartesian coordinates, 
k a combined index of reciprocal space vector $\vec k$ and band index $\nu$,
$\epsilon(k)$ the band energy,
and $v_\alpha(k)$ the group velocity in the direction $\alpha$.
To obtain the anisotropy \SigRatio as a point of reference, 
we assume in the vicinity of the band edges parabolic bands in an
anisotropic effective mass model.
The inverse mass tensor $\TENS M$ close to the band edges is
diagonalized in the form
\begin{eqnarray}
\TENS M&=& \TENS m^{-1} = \text{diag}(M_1,M_2,M_3), 
\quad\text{with eigenvectors}
\nonumber\\
\vec c_i&=& c_{i,x}\vec e_x +c_{i,y}\vec e_y+c_{i,z}\vec e_z
\quad (i=1 \dots 3)
\label{eq-mass-tensor-gen}
\qquad ,
\end{eqnarray}

with $\vec e_\alpha$ the basis vectors of the Cartesian coordinate
system.
The transport distribution for the anisotropic effective mass model
along the main axes $\vec c_i$ of the effective mass ellipsoid
are proportional to

\begin{eqnarray}
\sigma_{ii} &\propto& \frac{\sqrt{m_1 m_2 m_3}}{ m_i}
\\
                    &\propto& \frac{\bar{m}}{m_i} = \bar{m} M_i
\quad \text{with}
\nonumber\\
\bar m &=& \sqrt{m_1 m_2 m_3}
\qquad ,
\end{eqnarray}


where the masses have to be chosen positive for both the valence band
maximum and the conduction band minimum.

The following expressions for
the in-plane and cross-plane conductivities
project the mass tensor of general orientation given
by eq.~(\ref{eq-mass-tensor-gen}) to the Cartesian coordinates.
\begin{eqnarray}
\sigma_{xx} &\propto& \vec e_x^{\, T} \TENS M \vec e_x
\nonumber\\
            &\propto& (c_{1,x})^2 M_1 + (c_{2,x})^2 M_2 +(c_{3,x})^2 M_3
\nonumber\\
\sigma_{yy} &\propto& (c_{1,y})^2 M_1 + (c_{2,y})^2 M_2 +(c_{3,y})^2 M_3
\nonumber\\
\sigma_{zz} &\propto& (c_{1,z})^2 M_1 + (c_{2,z})^2 M_2 +(c_{3,z})^2 M_3
\label{eq-cip-cpp-conduct}
\qquad .
\end{eqnarray}

Due to the space group $D^5_{3d}$ ($R\bar{3}m$) the considered band extrema are 
two- and six-fold degenerate. In the case of the rhombohedral lattice
this summation leads to equal contributions $\sigma_{xx}$ and
$\sigma_{yy}$ and keeps the $\sigma_{zz}$ unchanged.
The in-plane transport distribution is given for symmetry reasons by 
$ \sigma_{\parallel} = \left(\sigma_{xx} + \sigma_{yy}\right)/2$.
and the cross-plane component by $\sigma_{\perp} = \sigma_{zz}$.
In the following the term $\sigma_{xx}$ will be used synonymously for
the in-plane component $\sigma_{\parallel}$, so the anisotropy is
denoted as \SigRatio.

The mass tensor in \BiTe is parameterized close to the band edges
using the calculated band structure
on a very dense mesh in k-space 
corresponding to 400 points along a reciprocal lattice vector
to obtain convergence concerning the
position of the extremum and the inverse mass tensor. The
values for the valence and the conduction band are summarized in
Tabs.~\ref{tab-bite-val}.~and \ref{tab-bite-cond}. \cite{yavorsky11}



\begin{table}[!h]
\begin{tabular}{|l|l|l|l|l|}
\hline
VBM   & $M_i$    & $c_{i,x}$  & $c_{i,y}$  & $c_{i,z}$   \\
\hline
 1 & -41.3  & 0.5000 & -0.8660 &  0.0000 \\
\hline
 2 & -7.42  & 0.6000 &  0.3463 &  0.7212  \\
\hline
 3 & -0.507 & 0.6246 &  0.3606 & -0.6927   \\
\hline
\end{tabular}
\caption{
Inverse effective mass tensor eigenvalues
$M_i$ and eigenvectors $\vec c_{i}$ at the
valence band maximum (VBM). 
}
\label{tab-bite-val}
\end{table}

The position $\vec q_S$ of the six-fold degenerate valence band maximum
in units of inverse Bohr radii is (0.372199,0.644655,-0.0299675) on
the plane ($\Gamma$ZU).
The effective mass ellipsoid is very anisotropic.
The angle $\phi$ of
the long axis of the ellipsoid with the (xy) basal plane is
43.8$^{\circ}$,
which is in good agreement to other calculations and experiments,
which show a quite spread.
\cite{mishra97,koehler76,stordeur88}
The transport anisotropy ratio determined by the effective mass tensor
for hole states close to the valence band maximum accounts to
$\SigRatio = 5.40$.


\begin{table}[!h]
\begin{tabular}{|l|l|l|l|l|}
\hline
CBM   & $M_i$    & $c_{i,x}$  & $c_{i,y}$  & $c_{i,z}$   \\
\hline
 1 & 5.62  & 1.     &   0.     & 0.  \\
\hline
 2 & 5.62  & 0.     &   1.     & 0.  \\
\hline
 3 & 1.20  & 0.     &   0.     & 1.   \\
\hline
\end{tabular}
\caption{
Inverse effective mass tensor eigenvalues
$M_i$ and eigenvectors $\vec c_{i}$
at the Conduction band minimum (CBM).
}
\label{tab-bite-cond}
\end{table}

Tabs.~\ref{tab-bite-cond} summarizes the effective mass tensor at the
two-fold degenerate conduction band minimum with a the position $\vec q_S$ 
in units of inverse Bohr radii (0,0,0.0552), which is about one third
of the line($\Gamma$Z).
The transport anisotropy is $\SigRatio = 4.67$.

With these values the convergence of the transport distribution
given by eq.~(\ref{eq:def-sigma})
determined by interpolation schemes in the Brillouin zone for
energies close to the band edges can be quantified.

Using the calculated band structures and the derived effective
mass tensors we obtain an anisotropy \SigRatio of 5.40 and 4.67 at the
valence and conduction band edge, respectively.
These values are marked by dots in Figs.~\ref{fig:bs-sig-max}) and
\ref{fig:bs-sig-min}).

\section{Conductivity anisotropy: comparison of different 
interpolation scheme}

\begin{figure*}[!htb]
\includegraphics[width=\PICTww]{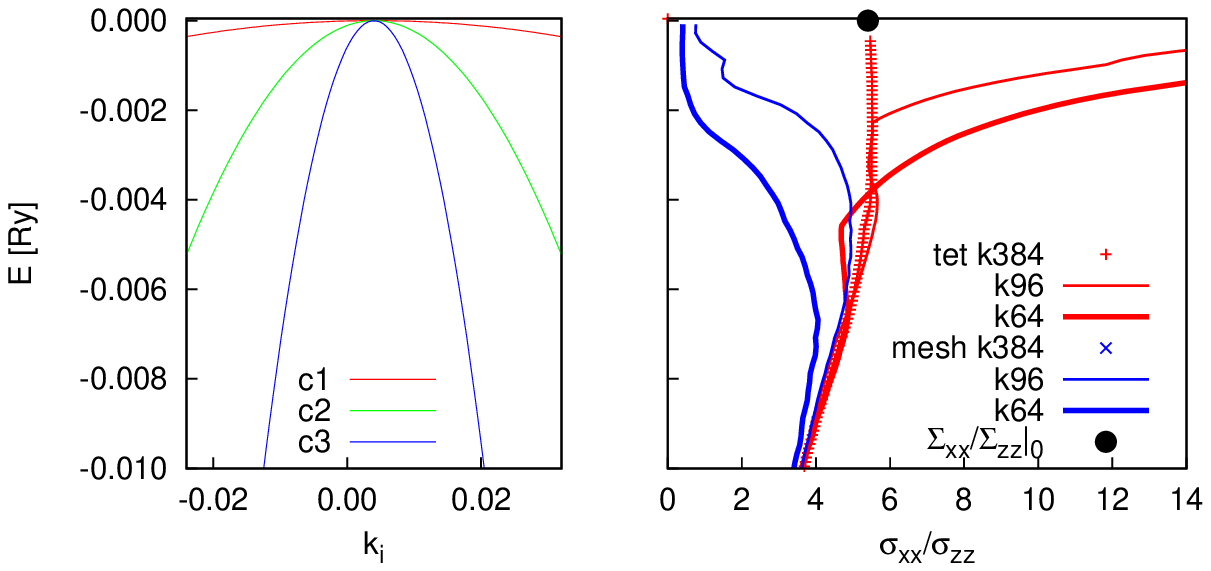}
\caption{Band structure of \BiTe (left) 
  and conductivity ratio (right) near valence
  band maximum, group velocity interpolation by
  tetrahedron method (tet) and mesh velocity concept (mesh),
  number of k-points along a reciprocal lattice vector is
  given as parameter. Observe that the vertical axis gives the energy
  and the anisotropy \SigRatio is on the horizontal axis.
\label{fig:bs-sig-max}
}
\end{figure*}

\begin{figure*}[!htb]
\includegraphics[width=\PICTww]{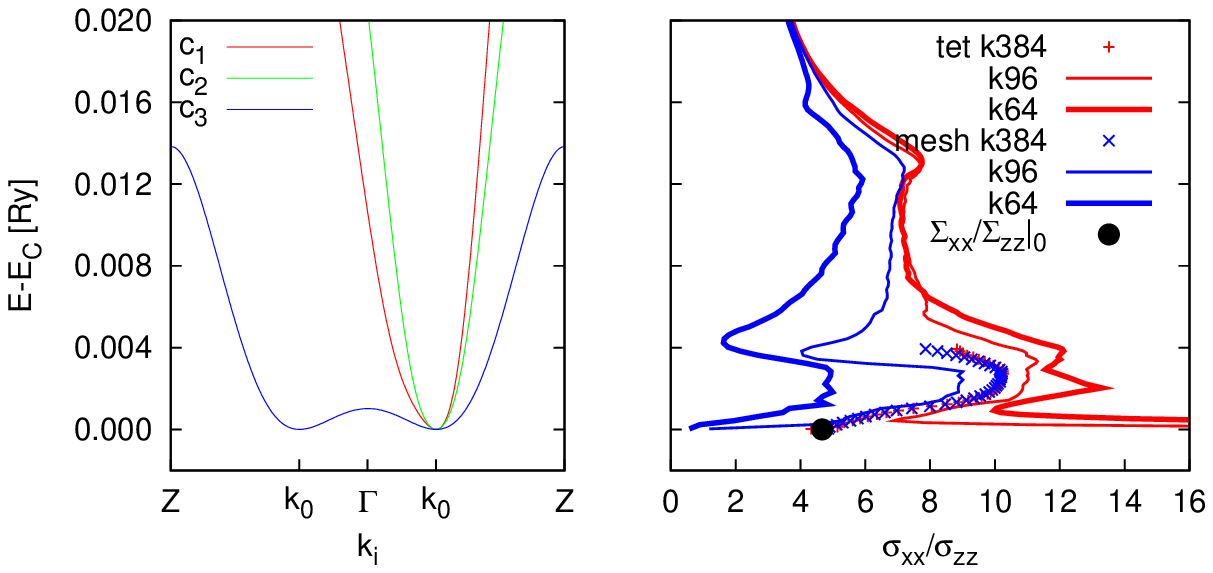}
\caption{Band structure of \BiTe (left) 
  and conductivity ratio (right) near conduction
  band minimum. see fig.~\ref{fig:bs-sig-max}
\label{fig:bs-sig-min}
}
\end{figure*}

The transport distribution $\sigma(E)$ of \BiTe is calculated by two methods.
The main distinction is the determination of the group velocities
$\vec v(k)$.

The tetrahedron method divides the irreducible part of the Brillouin
Zone (BZ) into disjoint tetrahedra. The group velocity is obtained by
a linear interpolation of the band energies at the 4 corner points and
approximates $\vec v(k)$ in the volume of the tetrahedron.
\cite{lehmann72}
The second method determines the velocities as derivatives along the
lines of the Bl{\"o}chl mesh
\cite{bloechl94}
in the whole Brillouin zone. The
directions of these lines are parallel to the reciprocal space vectors
and so the anisotropy of the real lattice is reflected in these
vectors. For \BiTe with a large ratio $c^{\text{hex}}/a^{\text{hex}}$
of 6.95 \cite{landolt-iii-41c} the real space unit cell is very
prolongated and the reciprocal lattice vectors are
quite close to the (xy) basal plane with an angle
$\theta'=118.05^{\circ}$ between them very close to the maximum value
of $120^{\circ}$. 
Projecting back these so-called 
mesh velocities to the Cartesian components quite large errors occur
in the resulting velocities as discussed below.
The results of the integration of eq.~(\ref{eq:def-sigma}) using both
interpolations schemes and different densities of k-points are compared
in Figs.~\ref{fig:bs-sig-max}) and ~\ref{fig:bs-sig-min}).

We start the discussion with the valence band maximum because the
principal axes of the mass tensor are directed along the x, y and z axis.
The left hand panel of Fig.~\ref{fig:bs-sig-max} shows the parabolic
dispersion along the three principal axes with the very different effective
masses, spreading by a factor of about 80. The right hand panel summarizes
the anisotropy ratios \SigRatio as a function of energy. 
It is obvious that both interpolation schemes give systematic
deviations from the expected values. For very large k-mesh densities
defined by 384 mesh points along a reciprocal lattice vector
the results converge to the correct value at the band edge. 
The necessary densities of
k-points are very demanding for realistic band structure calculations
with some atoms in the unit cell. 
The tetrahedron method overestimates systematically the
anisotropy ratio, whereas the mesh velocity method underestimates the
values. 
This behavior will be discussed below for a simpler two-dimensional
lattice and the found trends are confirmed.




A similar behavior is obtained for the conduction band minimum as
shown in Fig.~\ref{fig:bs-sig-min}). Here, a second challenge appears
with a local maximum of the conduction band at the $\Gamma$ point very
close in energy to the conduction band minimum. In addition, a large
transport anisotropy is obtained due to the occurrence of saddle points
in the band structure, as discussed elsewhere.\cite{yavorsky11}

The obtained anisotropies for the transport distribution can be
strongly influenced by the k-mesh density especially close to the band
edges.
Taking the converged transport distributions $\Sigma(E)$
we
obtained for small p-doping an anisotropy of about 6, and for
small n-doping an anisotropy of about 9. \cite{hinsche11a}
This about a factor of 2 larger than found in
experiment\cite{delves61} and other calculations.
\cite{scheidemantel03,huang08,park-freeman10}
Assuming an anisotropy of 0.47 for the averaged relaxation time of
states traveling 
along and across the basal plane we obtain a good agreement with
experiment. The reason for this scattering anisotropy has
to be elucidated by calculations of the microscopic transition
probabilities caused by defects or other scattering centers.

\section{Interpolation schemes: tetrahedron vs. mesh velocities}

\subsection{Mesh velocity method}

\begin{figure}[!htb]
\includegraphics[width=\PICTw]{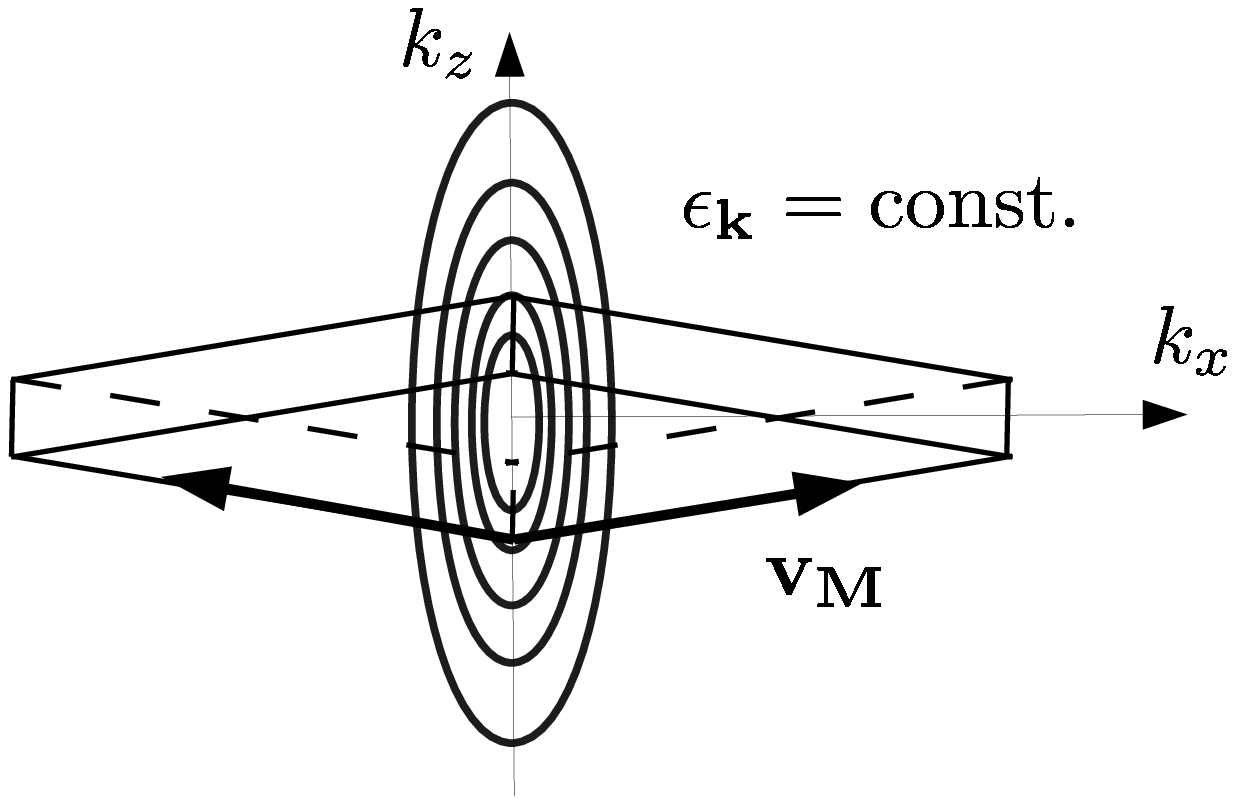}
\caption{Determination of the group velocity by means of so-called
  mesh velocities $v_M$ along the reciprocal basis vectors. 
  Anisotropic band dispersion is sketched by elliptical
  isoenergetic lines $\epsilon_k=\text{const.}$ 
  in the ($k_x$,$k_z$)-plane. 
  The rhombohedral Brillouin zone for the case of \BiTe 
  is shown by thin lines.
\label{fig:mesh-vel}
}
\end{figure}

\begin{figure}[!htb]
\includegraphics[width=\PICTw]{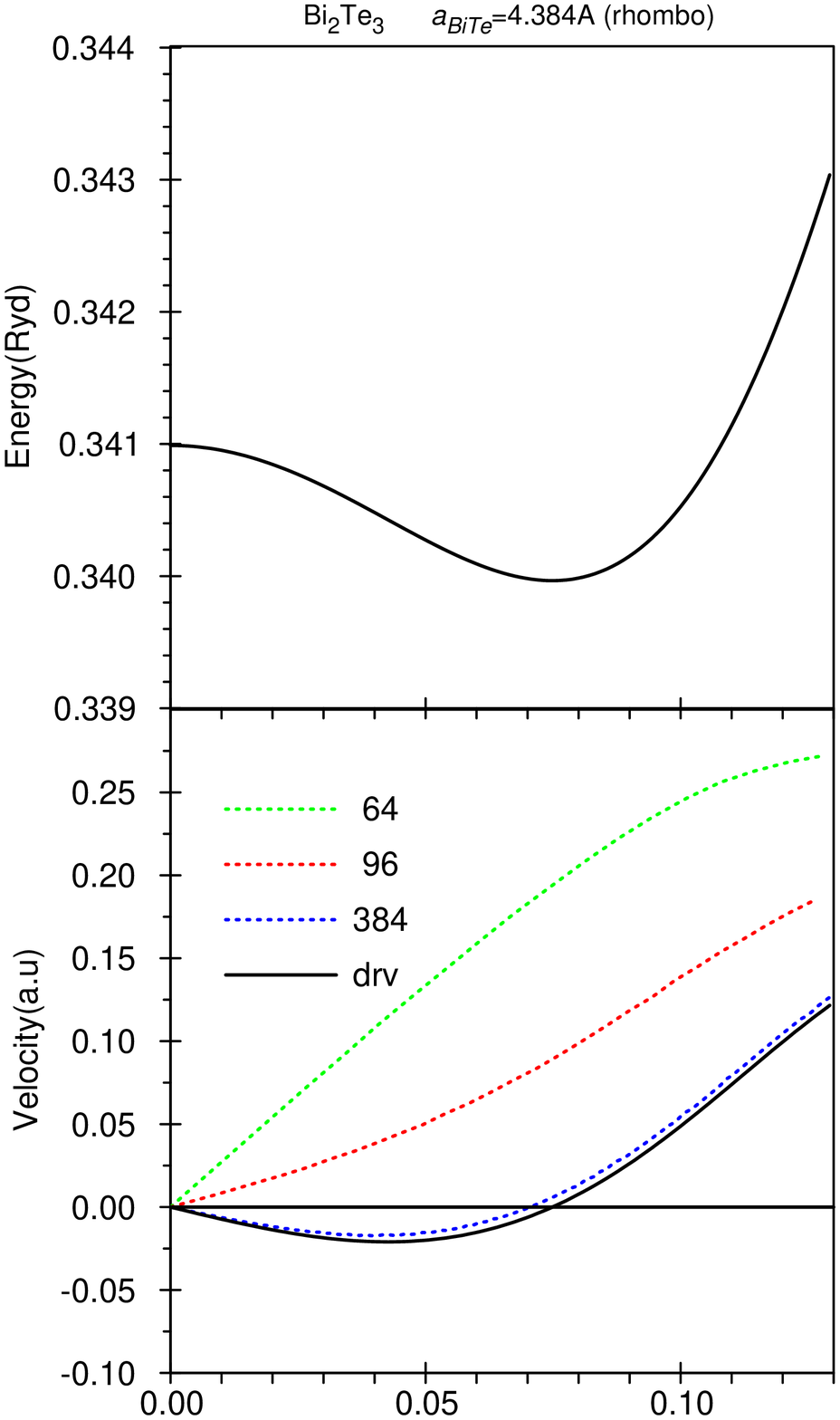}
\caption{Band structure $\epsilon(k_z)$ of \BiTe on 
  the line $\Gamma$-Z (top) and mesh velocities in 
  z direction (bottom) for different densities of the k-mesh. 
  'drv' denotes the exact result from the derivate
  $\partial\epsilon(k_z)/\partial k_z$.
\label{fig:mesh-vel2}
}
\end{figure}

In the following the implications of the different interpolation
schemes for the group velocity are discussed.
We start with the discussion of the so-called mesh velocities. They
are determined by a numerical derivative of the band dispersion along
the directions of the reciprocal lattice vectors which span the
Bl{\"o}chl mesh. The Cartesian components of the velocities are
obtained by a non-orthogonal transformation. If the angles between the
basis vectors are very large (the maximum is $120^{\circ}$) small
errors can be largely enhanced. This appears especially at band
structures with a strong anisotropy between directions in (xy)-plane
and the z-direction.

Fig.~\ref{fig:mesh-vel} shows schematically the band dispersion of
\BiTe close to the conduction band minimum by isoenergetic lines.
The Brillouin zone is shown in reduced size but
the angles correspond to the considered case of \BiTe. As it is
obvious, the interpolation of the velocities deviates strongly from
the correct values for k-points close to the band extremum. 
The directions of the reciprocal basis vector mainly scan along
the (xy)-plane. If the anisotropy of the band dispersion $\epsilon(k)$
is very strong as in the case considered, the mesh velocities tend to
equalize the in-plane and out-of-plane components of the velocity
$v_{x,y}\approx v_z$ which leads to an anisotropy closer to
unity. This can be seen in Figs.~\ref{fig:bs-sig-max}) and
~\ref{fig:bs-sig-min}) by the curves labeled 'mesh'. This effect is
most pronounced close to the band edges.

A quantitative comparison of
the accuracy of the mesh velocities in given in
fig.~\ref{fig:mesh-vel2}. 
To compare the mesh velocities with the exact values we have chosen
the $\Gamma$Z line and consider the z-component of the velocity
$v_z(k_z)$. 
The minimum of the energy dispersion is the global minimum of the
conduction band. A local maximum appears at the $\Gamma$ point.
For sparse k-mesh densities a strong deviation of the mesh velocities
have to be stated. To reproduce the point of inflection of the band
dispersion by the sign change in the velocity a very dense mesh is
necessary. 
The velocity in z-direction is overestimated by this method, if the
velocities in the (xy) directions are larger than in z-direction, as
illustrated in fig.~\ref{fig:mesh-vel} and typical for \BiTe in both
the valence and conduction band. As a result the transport anisotropy
\SigRatio, which is de facto the ratio of the velocities squared, is
shifted to unity. An increase of the anisotropy towards 1 is expected
for the opposite case with larger velocities along z-direction than
along (xy) directions.

\subsection{Tetrahedron method}

\begin{figure}[!htb]
\includegraphics[width=\PICTw]{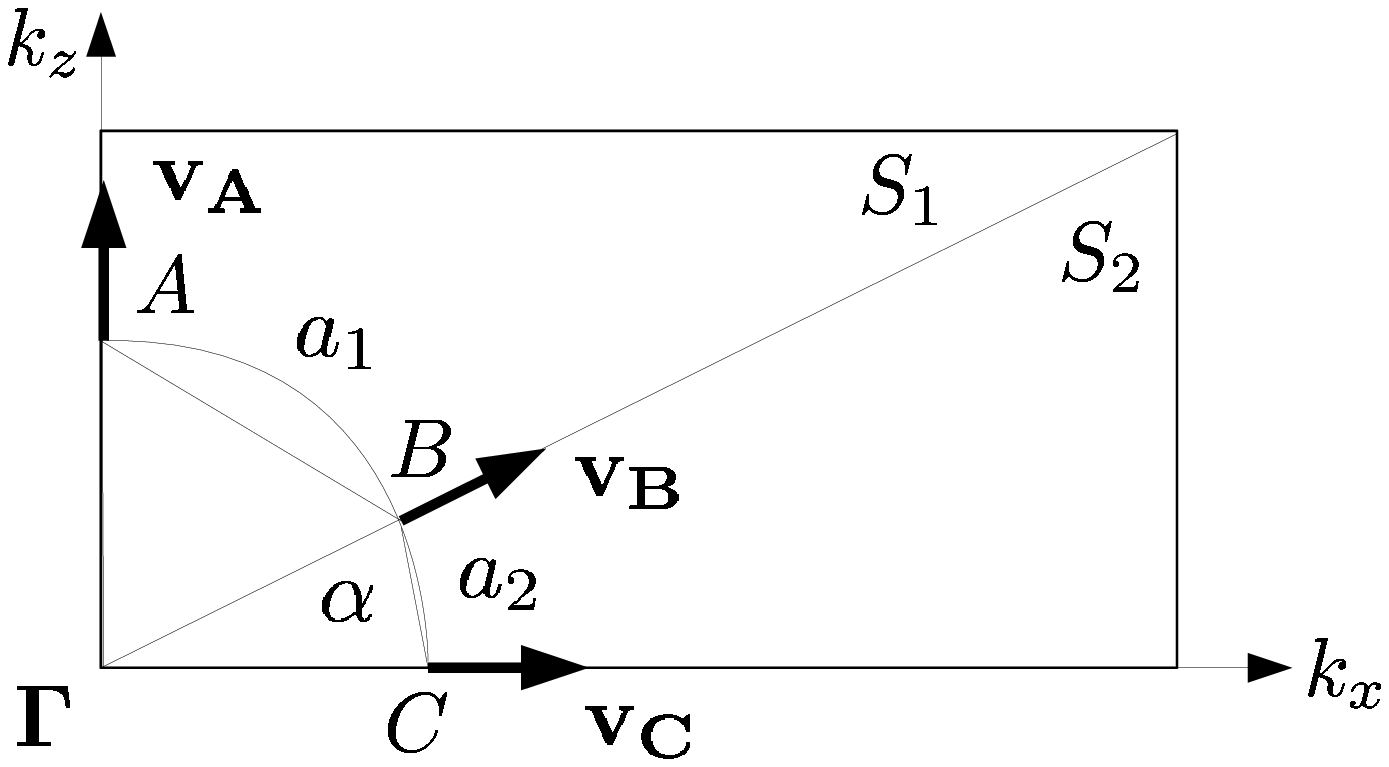}
\caption{Schematic sketch of an anisotropic k-mesh division
  characterized by the angle $\alpha$.
  The constant energy surface is approximated by the polygon ABC with
  the parts $a_1$ and $a_2$ in the triangles S$_1$ and S$_2$.
  $\vec v_A$, $\vec v_B$, and $\vec v_C$ denote the velocities.
  \label{fig:aniso}
}
\end{figure}

\begin{figure}[!htb]
\includegraphics[width=\PICTw]{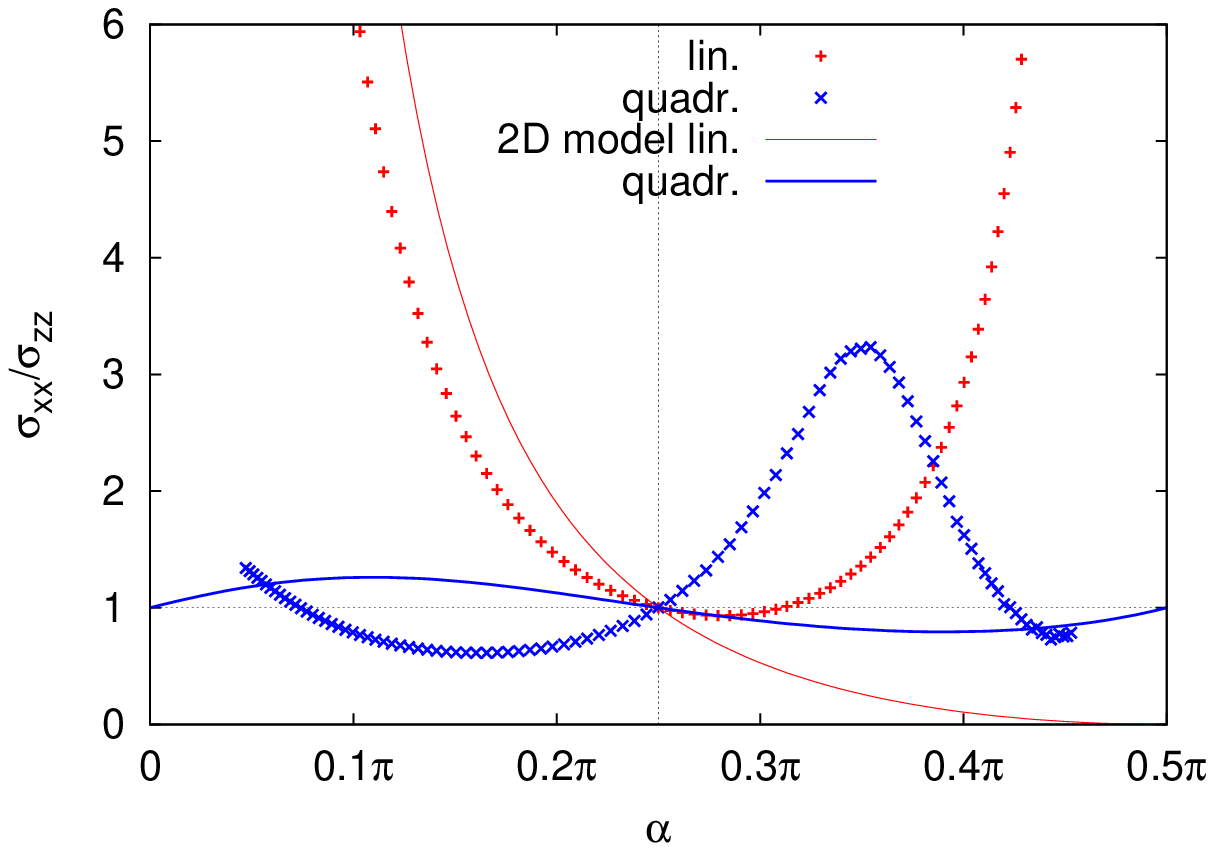}
\includegraphics[width=\PICTw]{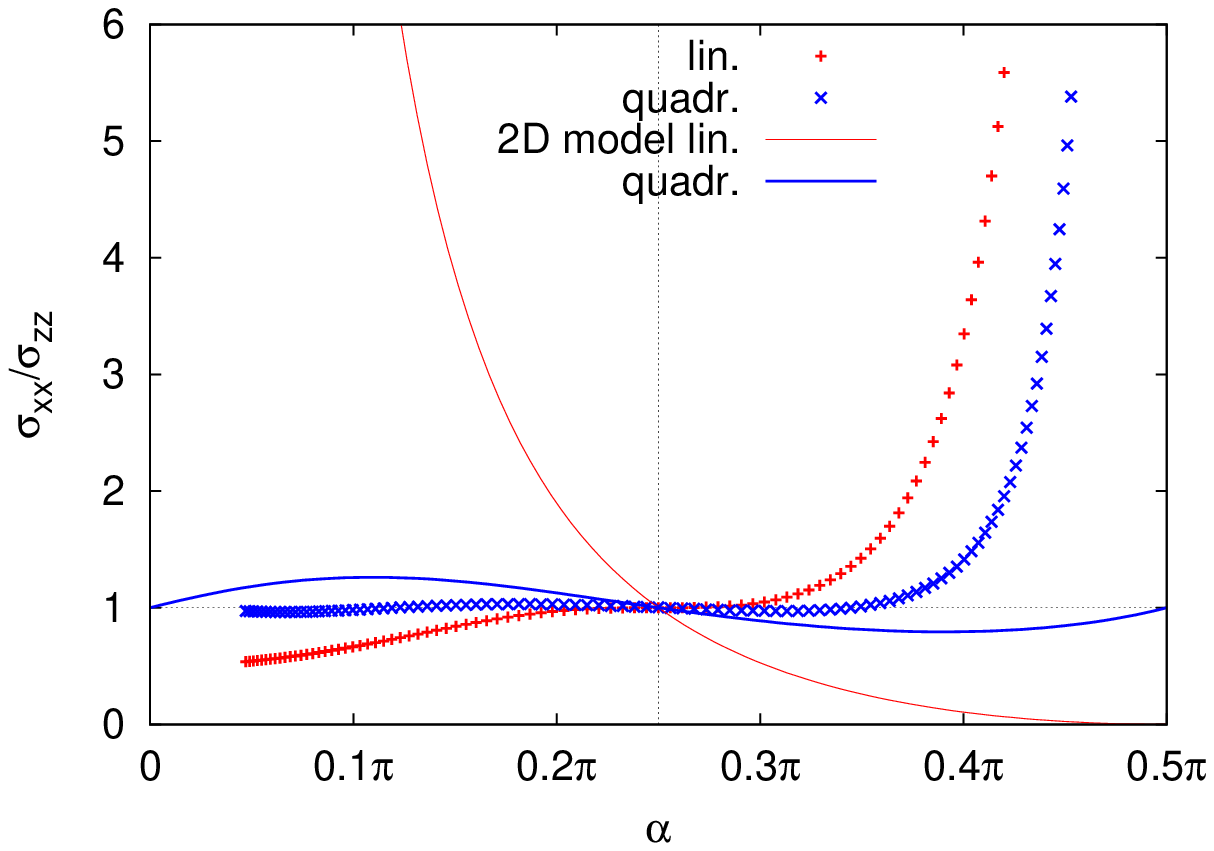}
\caption{Anisotropy of conductivity for a free electron model in
  dependence of the assumed underlying rhombohedral lattice
  characterized by the angle $\alpha$ between the reciprocal lattice
  vectors and the (xy) basal plane. 
  As comparison the conductivity anisotropies of the 2D model 
  (linear and quadratic interpolation) are
  shown with solid lines.
  Top panel: k-mesh created along ($11\bar{1}$) directions, 
  bottom panel: k-mesh created along (001) directions.
\label{fig:sig-ratio-tetra}
}
\end{figure}

The capability of the tetrahedron method to calculate the anisotropy of
the transport distribution will be evaluated for a two-dimensional
mesh for the cases of linear and quadratic interpolation of the
velocities. The results for a free electron dispersion with two
different arrangements of tetrahedrons in the irreducible part of a
three-dimensional meshes will be compared with.

The regular filling of the two-dimensional lattice is characterized by
the angle $\alpha$ between the border line of the triangles S$_1$ and
S$_2$ and the basal (xy)-plane, see Fig. \ref{fig:aniso}. 
Depending on the interpolation method
the velocities have to be determined for the triangles S$_1$ and
S$_2$ with a linear scheme and can be determined at the points A, B,
and C with a quadratic interpolation scheme. The second task requires
the knowledge of the energy dispersion $\epsilon(k)$ at additional
points on the edges.
Solving the linear set of equations for the Cartesian components of
the velocities $\vec v_1$ and $\vec v_2$ for a free electron
dispersion, one finds both velocities equal pointing in the same
direction along the borderline of the two triangles 
$\vec v_1 || \vec v_2|| (\cos\alpha,\sin\alpha)$. So, the anisotropy of
the transport distribution \SigRatio is just given by the ratio of the
components $v_x$ and $v_z$ squared

\begin{equation}
\SigRatio =
\frac{\cos^2\alpha}{\sin^2\alpha}
=
\frac{1}{\tan^2\alpha}
\label{eq:sig-rat-lin}
\end{equation}

For the given c/a ratio of \BiTe the angle $\alpha$ is about
23$^\circ$ which results an anisotropy value of 5.8 which is much to
large in comparison of the expected unity ratio.
Improving the interpolation scheme of the velocity to second order the
velocities 
$\vec v_A$, $\vec v_B$, and $\vec v_C$ 
are determined correctly and the error of the transport distribution
function is mainly determined by the approximation of the Fermi
surface (a line in the two-dimensional case) by the two line
segments $a_1$ and $a_2$. Due to the canceling of prefactors in the
ratio \SigRatio it is sufficient to consider the sum of the segment
lengths $|a_1|$ and $|a_2|$ times the square of 
the corresponding velocity components of 
$\vec v_A$ and $\vec v_B$ for area S$_1$, and  
$\vec v_B$ and $\vec v_C$ for area S$_2$
, see eq.~(\ref{eq:def-sigma})

\begin{eqnarray}
\sigma_{xx} &\propto&
\cos^2\alpha    \sqrt{(1-\sin^2\alpha)^2+\cos^2\alpha} +
\nonumber
\\
&& + (1+\cos^2\alpha)\sqrt{(1-\cos^2\alpha)^2+\sin^2\alpha}
\nonumber
\\
&\propto&
\cos^2\alpha    \sqrt{2-2\sin\alpha} +
(1+\cos^2\alpha)\sqrt{2-2\cos\alpha}
\nonumber
\\
\sigma_{zz} &\propto&
\sin^2\alpha    \sqrt{2-2\sin\alpha} +
(1+\sin^2\alpha)\sqrt{2-2\cos\alpha}
\quad .
\label{eq:sig-rat-quadr}
\end{eqnarray}

From these expressions the ration \SigRatio is evaluated and shown as
(thicker) blue lines in
fig.~\ref{fig:aniso} together with the values obtained by the
linear velocity interpolation as (thinner) red lines. 
The linear scheme shows strong
deviations especially for very anisotropic lattices with small or
large angles $\alpha$. This error is much reduced by the second order
interpolation scheme with a maximum error of about 30\%. 

Performing these procedures in a three-dimensional lattice requires
the set-up of k-point mesh filling the irreducible part of the Brillouin
zone and allowing for a disjunct tetrahedron arrangement. Two filling
schemes are evaluated, which are based on cubes (in a given basis)
which are filled with 6 tetrahedron each if completely inside the irreducible
part, otherwise they are partially used to fill the irreducible part.
The first one is based on cubes with the main
axes directed along the ($11\bar{1}$) reciprocal lattice directions.
The results for the anisotropy \SigRatio with a linear and a quadratic
velocity interpolation
are shown as crosses and plus signs in the upper panel of 
fig.~\ref{fig:sig-ratio-tetra}, respectively.
To define a similar angle $\angle$ characterizing the anisotropy of
the rhombohedral reciprocal lattice as in the two-dimensional case 
we have chosen 
$(\pi-\angle(\vec g_1,\vec g_2+\vec g_3))/2$ to ensure the isotropic
simple cubic lattice to be characterized by an angle $\pi/4$.

The second filling scheme uses the ($001$) directions as basis. The
results are shown in the lower panel of 
fig.~\ref{fig:sig-ratio-tetra}.
The results of the two-dimensional model can be partially reproduced,
especially the case of a isotropic lattice at an angle of
45$^\circ$. On average the deviations are larger for the linear
interpolation scheme in comparison to the quadratic one as 
expected from the two-dimensional model.
The partially opposite behaviour of the three-dimensional integration
scheme with respect to the two-dimensional one needs further
clarifications.

The results in fig.~\ref{fig:sig-ratio-tetra} are evaluated for
energies very close to the free electron band minimum. In these cases the
isoenergetic surface is approximated by a few trigonal elements only and the
largest anisotropies caused by the interpolation errors of velocities and
surface areas are expected. For larger energies these discrepancies
disappear quickly. The discussed deficiencies of the interpolation
schemes in k-space are restricted to energies close to band
extrema, which appear in transport properties of medium-doped
semiconductors. In cases of very small doping the application of an
anisotropic effective mass model seems to be more advantageous.

\section{Conclusions}

By means of model and realistic band structure calculations for
crystals with rhombohedral symmetry we have shown that the
determination of the transport distribution function requires very
dense meshes in k-space. Two different methods to determine the group
velocities are evaluated. It is found that they underestimate and
overestimate the transport anisotropy in a systematic manner.
For very prolongated unit cells the anisotropy in k-space requires a
thorough check of the convergence with respect to k-space density.

\bibliography{all_lit}

\end{document}